\documentclass[%
 aip,
 amsmath,amssymb,
 reprint,%
 nofootinbib
]{revtex4-1}

\usepackage{graphicx}
\usepackage{dcolumn}
\usepackage{bm}
\usepackage{enumerate}

\usepackage{setspace,color,url}
\allowdisplaybreaks[1]

\newcommand{\kt}{k_{\text{B}}T}
\newcommand{\UU}{{\cal U}}

\usepackage[dvipsnames]{xcolor}
\newcommand\revr[1]{\textcolor{black}{{#1}}}

\begin{document}

\preprint{Submitted to The Journal of Chemical Physics}

\title{Osmotic and diffusio-osmotic flow generation at high solute
concentration. \\ I. 
{Mechanical} approaches}
%
\author{Sophie Marbach}
\email{sophie.marbach@lps.ens.fr}
\affiliation{LPS, UMR CNRS 8550, Ecole Normale Sup\'erieure, PSL Research University, 24 rue
Lhomond, 75005 Paris, France}
\author{Hiroaki Yoshida}
\email{h-yoshida@mosk.tytlabs.co.jp}
\affiliation{LPS, UMR CNRS 8550, Ecole Normale Sup\'erieure, PSL Research University, 24 rue
Lhomond, 75005 Paris, France}
\affiliation{Toyota Central R\&D Labs., Inc., Nagakute, Aichi 480-1192, Japan}
\author{Lyd\'eric Bocquet}
\email{lyderic.bocquet@lps.ens.fr}
\affiliation{LPS, UMR CNRS 8550, Ecole Normale Sup\'erieure, PSL Research University, 24 rue
Lhomond, 75005 Paris, France}
%
\date{\today}
%
\begin{abstract}
In this paper, we explore various forms of osmotic transport in the regime of high solute concentration. 
We consider both the osmosis across membranes and diffusio-osmosis at solid interfaces, driven by solute 
concentration gradients. 
We follow a mechanical point of view of osmotic transport, which allows us to gain much insight into the
local mechanical balance underlying osmosis. 
We demonstrate in particular how the general  expression of the osmotic pressure for mixtures, as 
obtained classically from the thermodynamic framework, emerges from the mechanical balance controlling 
non-equilibrium transport under solute gradients. Expressions for the rejection coefficient of osmosis and the diffusio-osmotic 
mobilities are accordingly obtained. These results generalize existing ones in the dilute solute regime to mixtures with
arbitrary concentrations. 
\end{abstract}

\maketitle


\section{Introduction}

\begin{figure}[t]
\includegraphics*[width=0.8\columnwidth]{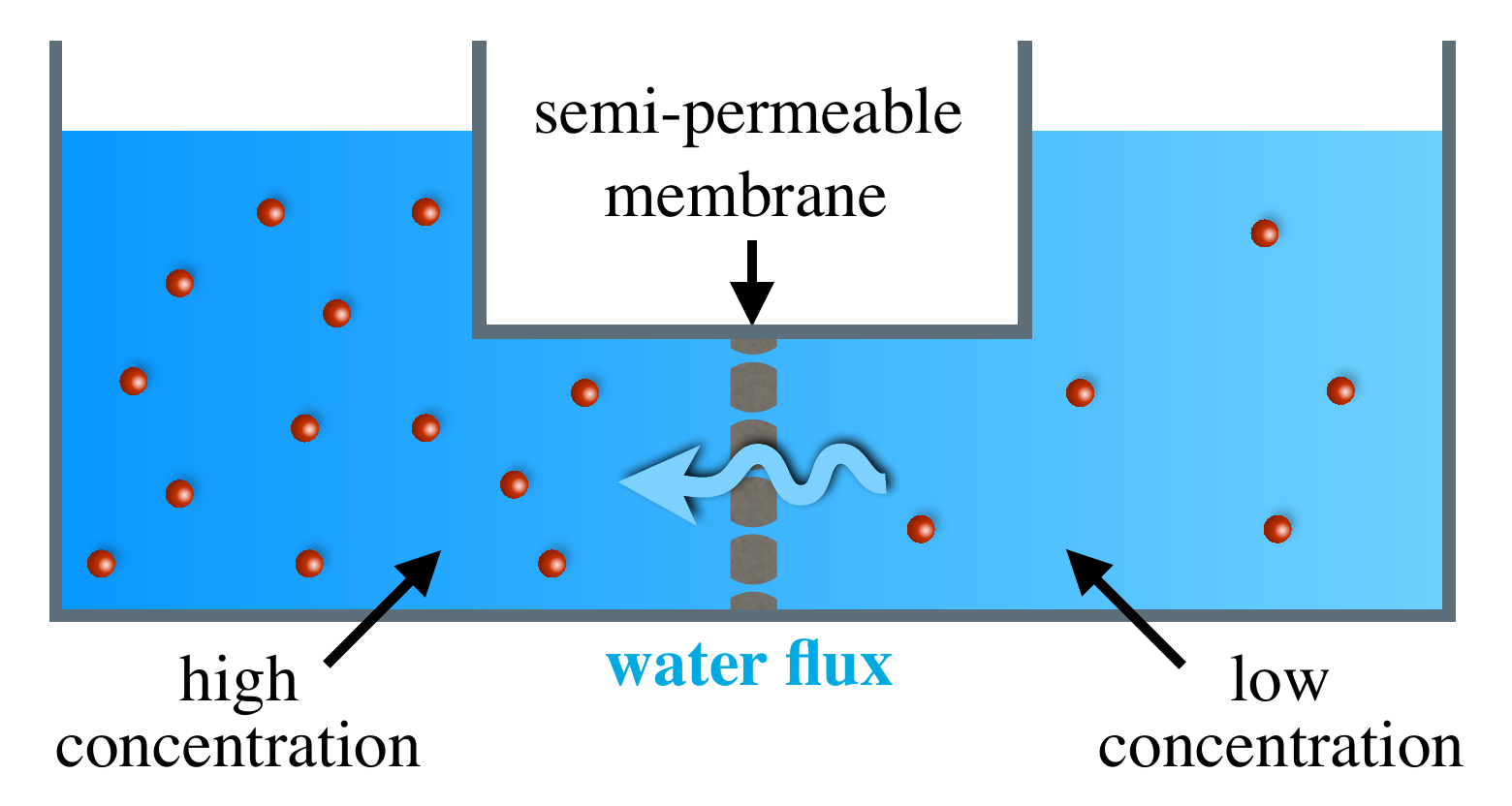}%
\caption{\label{fig:osmose} Geometry of osmosis. A semi-permeable membrane allows transport of water upon a solute concentration difference.
}
\end{figure}

Osmotic transport is a subtle and non-trivial effect, that is harvested in numerous biological phenomena and applications, such as food processing in biological organisms,~\cite{SB1987,JRH+2009} reverse osmosis for desalination, and energy generation from salinity differences,~\cite{KK1961,LE2012,SPB+2013,WOE2016} to name a few. Traditionally, osmotic transport is described as occurring across a semi-permeable membrane, {\it i.e.}, a membrane impermeable to the solute but permeable to the solvent, often water, see Fig.~\ref{fig:osmose}. If two reservoirs with different solute concentrations are put in contact via a semi-permeable membrane, an osmotic pressure builds up between the compartments. This pressure drop is the driving force for a flux of water from the low concentration reservoir 
to the highly concentrated one, until the thermodynamic equilibrium is reached. For low solute concentrations, the osmotic pressure is expressed by the van\;'t Hoff law, 
\begin{equation}
\Delta \Pi = \kt \Delta c, 
\end{equation}
where $\Delta c$ is the difference in solute concentration between the two reservoirs.~\cite{KK1961}
\revr{The van\;'t Hoff law is derived by equating the solvent chemical potential of the solvent across the membrane.~\cite{Gibbs1897,Guggenheim1985,KR2008} The osmotic pressure is accordingly defined in terms of equilibrium thermodynamic properties of the system.}
An interesting, and quite counterintuitive, remark is that -- provided it is semi-permeable -- the membrane characteristics do not appear in this
thermodynamic expression for the osmotic pressure.
Now, when the membrane is only {\it partially} impermeable to the solute, there is still a solvent flux driven
by the solute concentration imbalance.~\cite{TS1965A,*TS1965B,LCB+2014,LCF+2017}
However the driving osmotic pressure is usually assumed to be reduced by a (dimensionless) rejection factor, say $\sigma$. Determining $\sigma$ requires to describe the detailed mass and solute transport across the membrane, and this requires to go beyond the
thermodynamic description.
From a general perspective,
transport across a membrane is characterized within the framework of irreversible processes, via a transport matrix ${\mathbb L}$, relating fluxes to thermodynamic forces~\cite{KK1961,Manning1968,AB2006}
\begin{equation}
\label{Larray}
\left(\begin{array}{c} Q \\ \revr{J_s-cQ}\end{array}\right)= 
{\mathbb L}
\times \left(\begin{array}{c} - \nabla p \\ - \nabla \mu\end{array}\right),
\end{equation}
with $Q$ and $J_s$ denoting respectively \revr{the volume flux (per unit area) of the solution} and of the solute through the membrane;
$c$ is the solute concentration, $p$ is the pressure, and $\mu$ is the solute chemical potential. This matrix is symmetric according to Onsager's principle.
The question then amounts to characterizing the coefficients of this matrix
associated with osmotic gradients. 
Kedem and Kachalsky rewrote these transport equations in a more explicit form as~\cite{KK1961,KK1963A,*KK1963B,*KK1963C}
\begin{eqnarray}
&Q= -\mathcal{L}_\text{hyd} \left( \Delta p - \sigma \kt \Delta c\right), \label{kk1}\\
&J_s= - \mathcal{L}_\text{D} \omega \Delta c + c (1-\sigma) Q, \label{kk2}
\end{eqnarray}
where $\mathcal{L}_\text{hyd}=\revr{\kappa/(\eta L)}$ is the solvent permeance
\revr{with $\kappa$ the permeability (in units of a length squared)}, $\eta$ the fluid viscosity, and
$L$ the membrane thickness,
and $\mathcal{L}_\text{D}=D/L$ is the solute permeability with $D$ the diffusion coefficient of the solute. 
\revr{The Onsager symmetry relations for Eq.~\eqref{Larray} can be verified by exploring two limiting cases: the situation where $\Delta p = 0$  yields $Q=\sigma \mathcal{L}_\text{hyd}c\Delta\mu$ (using  $\Delta \mu=\kt \Delta c/c$ in the dilute case); and the situation where $\Delta c=0$ yields $J_s-cQ=\sigma\mathcal{L}_\text{hyd}c\Delta p$, as expected.}
The Kedem--Kachalsky result introduces the reflection coefficient $\sigma$ mentioned previously, that is dependent in particular on the relative permeability of the membrane to the solvent and the solute.~\cite{TS1965A,AM1974,*TS1965B} Interestingly, the non-dimensional coefficients $\omega$ and $\sigma$ are expected to be linearly related,~\cite{KK1961} as $1-\sigma \propto \omega$, a result that we will recover below.
Note that the previous Kedem--Kachalsky equations are valid in the regime of dilute solute concentration, where the van\;'t Hoff relationship applies. Generalizing them to mixtures with arbitrary volume fractions requires to introduce the general thermodynamic expression for the osmotic pressure and its link to non-equilibrium transport remains to be developed. 

In this paper, our goal is to get some insight into the physical principles of osmosis, while exploring the high solute concentration regime. We will make use of 
a mechanical approach to osmosis, which is particularly illuminating to identify the force balance underlying the osmotic phenomenon.
We will consider the two situations of bare osmosis across membranes and diffusio-osmosis at solid interfaces.
Osmosis is expected to occur under a solute imbalance across a semi-permeable membrane, which is permeable to the solvent but not to the solute.
In contrast, diffusio-osmosis is a surface-driven flow occurring under solute gradients. This form of transport has attracted increasing attention in the context of recent developments in micro- and nano-fluidic systems.~\cite{ACY+2008,YZP+2012,LCB+2014,SUS+2016,SNA+2016,LCF+2017}

\begin{figure}[t]
\includegraphics*[width=0.9\columnwidth]{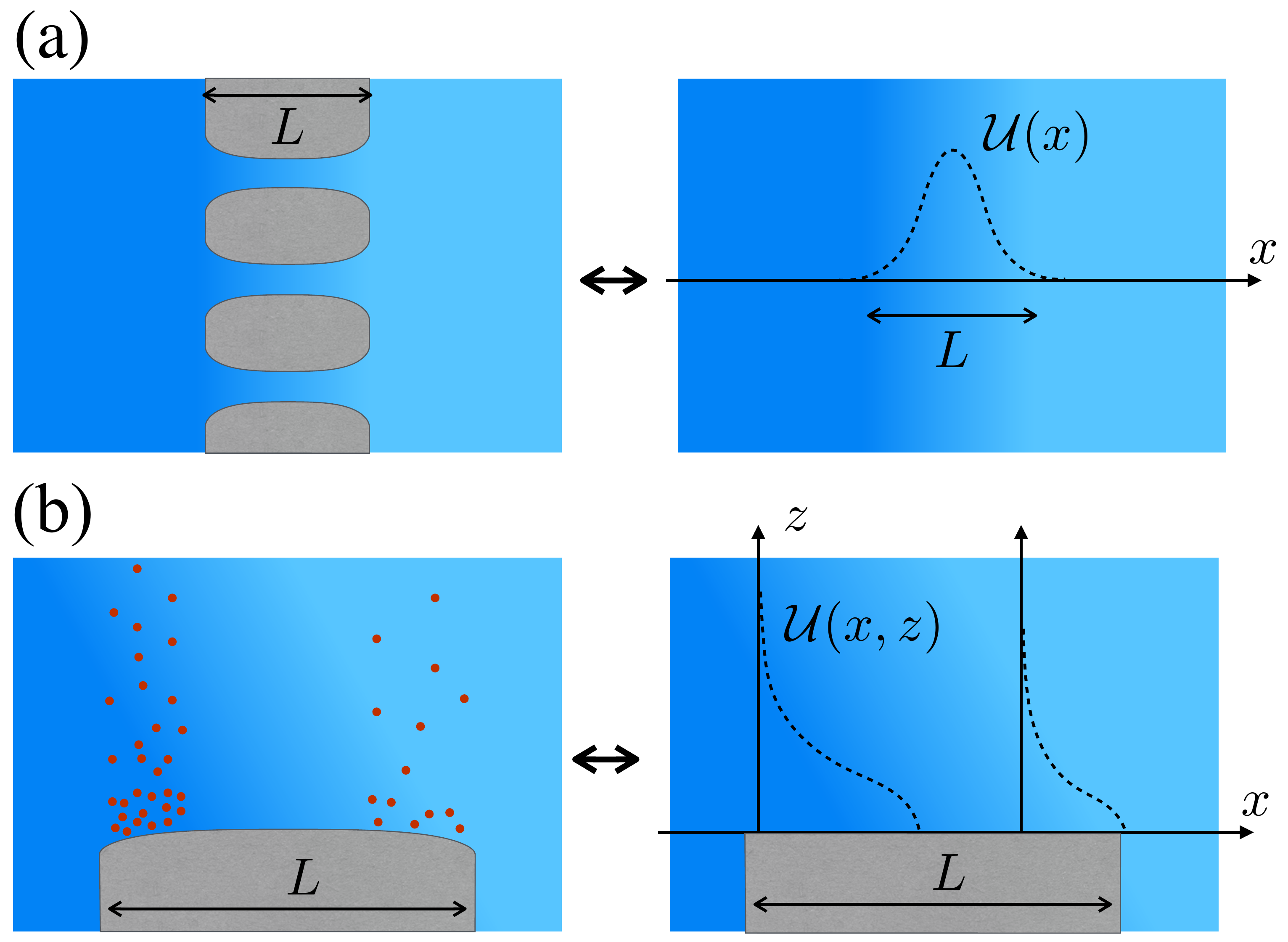}%
\caption{\label{fig:geo} From the hydrodynamic picture to the mechanical
 approach. (a) Geometry of osmosis: from the membrane type approach to
 the mechanical approach describing the membrane as a potential on the solute. (b) Geometry of diffusio-osmosis: from a zoomed surface-solute interaction to a description with a potential on the solute.
}
\end{figure}

To highlight the mechanical balance underlying osmotic transport, we will consider a simplified model where the effect of the membrane on the solute is described in terms of an energy barrier, see Fig.~\ref{fig:geo}(a).  
Such an energy barrier is a crude but convenient description for the membrane, avoiding to enter into the details of the interaction of the solute with the membrane. It reduces the description to its minimal ingredients of partial or semi-permeability, and makes it amenable to explicit calculations. 
As we show below, it allows to explore in details how osmotic pressure builds up, even in the absence of a full semi-permeability of the membrane to the solute.
{We note furthermore that such  barrier potential can also be physically achieved; for example, it can be generated from a nonuniform electric field acting on a polar solute in a nonpolar solvent,~\cite{Debye1954} or it can represent the nonequivalent interactions of solute and of solvent particles with a permeable membrane, {\it e.g.}, charge interactions.~\cite{GS1957,PGJ+2013} } 
In the case of osmosis, this approach was first introduced by Manning~\cite{Manning1968} in the low concentration regime, and generalized more recently by Picallo {\it et al.}  to explore the osmotic transport across perm-selective charged nanopores.~\cite{PGJ+2013}

On the basis of this mechanical approach, a further objective of our study is to explore osmotic phenomena in  the regime of high solute concentration, where the ``solute'' and the ``solvent'' are two components of a mixture with arbitrary molar fraction. 
In this case, thermodynamics predicts that the thermodynamic force driving motion is a generalized osmotic pressure
taking the formal expression:~\cite{BH2003}
\begin{equation}
\Pi(c)=c {\partial f\over \partial c} - f[c] + f[c=0],
\label{PiGeneral}
\end{equation}
with $f$ the free energy density (see Appendix A for a reminder). 
However, how this osmotic pressure is expressed in terms of mechanical balance across a membrane (osmosis) or along an interface (diffusio-osmosis)
has not been explored up to now. 
Our goal in the present work is accordingly to highlight how this thermodynamic expression connects to the (local) mechanical balance
for osmosis and diffusio-osmosis.

\section{\label{sec_osm}From a thermodynamic to a mechanical approach to osmosis}

We consider a membrane separating two sub-volumes, containing a solvent and a solute. The concentration difference between the two volumes is $\Delta c=c_+-c_-$. As introduced above, we assume that 
the membrane  behaves as an external potential $\UU$ on the solute {\it only}, but not on the solvent molecules. It varies only along the $x$ axis. 
We denote $L$ the lateral range of the potential $\UU$, so that $\UU(-L/2)=\UU(L/2)=0$, and vanishes outside of this domain, see Fig.~\ref{fig:geo}(a). 
Still we consider that the membrane is permeable to the solvent, with a permeance $\mathcal{L}_\text{hyd}$, relating the flux $Q$ to the pressure
drop $\Delta p$ in the absence of a concentration difference: $Q=-\mathcal{L}_\text{hyd} \Delta p$.

We first recall results for the dilute solute regime and then extend the results to the high concentration case.
%
\subsection{\label{sec_dilute}Dilute solute concentration}
Before considering the general case of high solute concentration, we first revisit the case of the dilute solute concentration, as explored in Refs.~\citenum{Manning1968,PGJ+2013}, which allows us to give the flavor of the approach. 
In the 1D geometry described above, the stationary solute concentration $c(x)$ obeys a Smoluchowski equation:
\begin{align}
0={\partial_t} c =& -\partial_x J_s  \nonumber\\
=& -\partial_x\left( -D \partial_x c + \lambda c \,(-\partial_x \UU) + v c \right),
\label{Smolu}
\end{align}
where $D$ is the diffusion coefficient and  $\lambda=D/\kt$ the mobility, with $k_{\mathrm{B}}$ and $T$ being the Boltzmann constant and the temperature, respectively.
We further assume a low P\'eclet number limit, $\mathrm{Pe}=v L/D \ll 1$, such that the convective term of Eq.~\eqref{Smolu} is negligible. This is valid for low permeability (nanoporous) membranes.
Since the solute current across the membrane $J_s$ is constant in time and spatially uniform, Eq.~\eqref{Smolu} is explicitly solved with respect to the concentration:
\begin{equation}
c(x)= -\Delta c\, e^{-\beta\UU(x)} { \int_{x}^{L/2} dx^\prime\, \exp[+\beta\UU(x^\prime)]\over \int_{-L/2}^{L/2} dx^\prime\, \exp[+\beta\UU(x^\prime)]},
\label{profile}
\end{equation}
where $\beta=1/\kt$.

Now let us focus on the force balance. It is crucial to remark that the membrane will act on the fluid as an external force, $-\partial_x \UU$, exerted on the solute molecules. But due to action-reaction, this force acts on the fluid volume on its globality. This is for example highlighted in the force balance on the fluid,
as represented by the  Stokes equation along the $x$ direction:
\begin{equation}
0 = -\partial_x p + c(x) (-\partial_x \UU)+ \eta \nabla^2 v,
\label{St}
\end{equation}
where 
$p$ is the fluid pressure and $v$ is the flow velocity of the fluid in the $x$ direction.
The driving force inducing the solvent flow is accordingly written in terms of an apparent pressure drop, $-\partial_x \mathcal{P}=-\partial_x p+c(x)(-\partial_x\UU)$. 
The membrane, via its potential $\UU$, will therefore create an average force on the fluid, which writes per unit surface 
\begin{equation}\label{eq_Fc}
-\Delta\mathcal{P}=-\Delta p+ \int_{-L/2}^{L/2} dx\, c \,(-\partial_x \UU),
\end{equation}
%
where $\Delta$ means the difference of quantity difference between two sides.
The second term of Eq.~\eqref{eq_Fc} can be interpreted as the osmotic contribution; 
one can calculate it explicitly 
using the concentration profile given in Eq.~\eqref{profile}, to obtain%
\begin{equation}
-\Delta \mathcal{P}= -\Delta p +\sigma \Delta \Pi.
\label{Starling}
\end{equation}
This leads to the classical van\;'t Hoff law of the osmotic pressure, $\Delta\Pi=\kt\Delta c$, and the expression of the reflection coefficient $\sigma$
is obtained as
\begin{equation}
\sigma= 1-{L \over \int_{-L/2}^{L/2} dx^\prime\, \exp[+\beta\UU(x^\prime)]}.
\label{reflectDilute}
\end{equation}
Equation~\eqref{Starling} is often referred to as the Starling equation in the physiology literature, see {\it e.g.} Ref.~\citenum{Pappenheimer1953}.

The above result correctly recovers the case of a completely semi-permeable membrane (no solute flux across the membrane), {\it i.e.}, $\beta \UU\gg 1$ in this limit, and thus $\sigma\rightarrow 1$, yielding $-\Delta \mathcal{P}=-\Delta [p-\Pi]$. 
In the intermediate cases, although the membrane is permeable, a flow arises due to the solute concentration gradient even in the absence of a pressure gradient. When the potential is repulsive and small $\UU \sim \kt$, then $0<\sigma<1$; the flow is in the direction of increasing concentration. When the potential is attractive, then $\sigma < 0$ and the flow reverses. 
\revr{Integrating Eq.~\eqref{St} over the membrane area ($\mathcal{A}$) and thickness ($L$) allows us to express the
total flux $Q$ as:}
\begin{equation}
Q = - \mathcal{L}_\text{hyd} \left(\Delta p -\sigma \kt\Delta c \right).
\label{KKbis}
\end{equation}
\revr{Here one may formally define the permeability $\kappa$ in terms of the averaged flow as $(1/\mathcal{A}L)\int\int dx\,d\mathcal{A}\,\nabla^2 v \equiv - {Q}/{\kappa}$, and the corresponding permeance $\mathcal{L}_\text{hyd} = \kappa/(\eta L)$. These parameters, $\kappa$ and $\mathcal{L}_\text{hyd}$, take into account  the detailed geometric specificities of the pores in the membrane. 
Overall Eq.~\eqref{KKbis} agrees with the Kedem--Kachalsky result in Eq.~\eqref{kk1}.}

We recall that according to Ref.~\citenum{KK1961} the 
reflection coefficient $\sigma$ and the factor $\omega$ are linked by a linear relationship in the form $1-\sigma \propto \omega$, for diffusion through the membrane. The origin of this symmetry relationship
is easily apparent from the general expression of the solute flux $J_s$. 
From the steady state condition of Eq.~\eqref{Smolu}, the solute flux is expressed as
\begin{equation}
J_s=- \lambda c\, \partial_x \left( \mu_{\rm id}[c]+{\cal U}\right),
\label{JsSemiGeneral}
\end{equation}
where \revr{$\mu_{\rm id}[c]=\kt\ln (c/c^*)$} is the chemical potential of the \revr{ideal (dilute)} solution with solute concentration $c$,
with $c^*$ being a reference concentration.
The first term can be rewritten as $- \lambda \partial_x \Pi$, with $\Pi=\kt c$.
The flux $J_s$ is spatially homogeneous ($\partial_x J_s=0$), so that one
deduces
\begin{align}
J_s &= -\lambda \left( \dfrac{\Delta \Pi}{L} - {1 \over L} \int_{-L/2}^{L/2} dx\, c(x) (-\partial_x {\cal U}) \right) \nonumber \\
&= - \dfrac{\lambda}{L} \left(1 - \sigma \right) \Delta \Pi,
\label{linkb}
\end{align}
and in the present case, $1-\sigma=\omega$.
Note that the contribution of the total flux $Q$ to the solute flux $J_s$ is recovered when the convective term of Eq.~\eqref{Smolu} is accounted for.~\cite{Manning1968}

%
\subsection{\label{sec_high}High solute concentration}

We now generalize the mechanical approach to the case of a mixture with a high solute concentration. 
As stated earlier, the osmotic pressure is expected in this regime to deviate from its van\;'t Hoff limit $\Pi=\kt c$, and is now defined in terms of the general thermodynamic expression given in Eq.~\eqref{PiGeneral} (as recalled in Appendix A.)~\cite{BH2003}

In this regime, the solute flux $J_s$ entering the Smoluchowski equation for the solute now writes
\begin{equation}
J_s=- \lambda[c(x)]\, c(x) \partial_x\left( \mu[c(x)] + {\cal U}(x)\right),
\label{JsDense}
\end{equation}
where $\mu[c]$ is the chemical potential of the solute and $\lambda[c]$ the solute mobility, possibly depending on the concentration.
The fluid equation of motion remains similar as above, in Eq.~\eqref{St}, with the membrane acting on the fluid in the
form of an external force $c(x)(-\partial_x{\cal U})$, leading to an average force as in Eq.~\eqref{eq_Fc}.

At equilibrium, fluxes are vanishing and the equilibrium concentration 
$c_0(x)$ thus obeys
\begin{equation}
\mu[c_0(x)]+{\cal U}(x)=\mu_{\rm res},
\end{equation}
where $c_0(-L/2)=c_0(L/2)=c_{\rm res}$, such that $\mu[c_{\rm res}]=\mu_{\rm res}$. 
Now when there is a concentration difference $\delta c_0$ of the solute between the reservoirs, solvent and solute fluxes build up. 
In contrast to the dilute case above, one cannot solve exactly the previous equations for $c(x)$. 
However, one may explore the case of a small concentration difference
between the reservoirs and compute the perturbation from equilibrium. 
This leads to a change in the concentration profile $c(x)$, which we write as $c(x)=c_0(x)+\delta c(x)$. At the boundaries, one has $\delta c(-L/2)=0$ and $\delta c(L/2)=\delta c_0$.
Equivalently, this can be expressed in terms of a chemical potential difference of the solute between the reservoirs, 
$\Delta \mu= \mu'[c_{\rm res}]\times \delta c_0$, where $\mu'=\partial \mu/\partial c$.

To lowest order in $\delta c(x)$, the flux in Eq.~\eqref{JsDense} now writes
\begin{equation}
J_s= -\lambda[c_0(x)] \, c_0(x) \partial_x\left( {\partial \mu\over \partial c}[c_0(x)]\, \delta c (x)\right).
\end{equation}
Since $\partial_x J_s=0$ in the stationary state, this equation is solved with respect to $\delta c(x)$:
\begin{equation}
\delta c(x)= - {J_s \over \mu^\prime[c_0(x)]} \int_{-L/2}^x {dx^\prime \over \lambda[ c_0(x^\prime)]\, c_0(x^\prime)}.
\end{equation}
Using the boundary condition $\delta c(L/2)= \delta c_0$,
we obtain the following expression for the flux:
\begin{equation}
J_s =  -{1 \over \int_{-L/2}^{L/2} {dx^\prime \over \lambda[ c_0(x^\prime)]\, c_0(x^\prime)} } \Delta \mu, 
\end{equation}
and deduce the concentration profile $\delta c(x)$ as
\begin{equation}
\label{deltaC}
\delta c(x) = {\Delta \mu \over \mu^\prime[c_0(x)]}  { {\int_{-L/2}^x {dx^\prime \over \lambda[ c_0(x^\prime)]\, c_0(x^\prime) }}
\over {\int_{-L/2}^{L/2} {dx^\prime \over \lambda[ c_0(x^\prime)]\, c_0(x^\prime)}}}.
\end{equation}
Using this solution for the density profile, we can now
compute the corresponding driving force due to the solute acting on the fluid, according to Eq.~\eqref{eq_Fc}:
\begin{align}
\int_{-L/2}^{L/2} dx\; \delta c(x)(-\partial_x{\cal U}) &=  c_0(L/2) \mu^\prime[c_0(L/2)] \delta c(L/2) \nonumber \\
&\hspace*{-2.5cm}- \mu^\prime[c_0(L/2)] \delta c(L/2) 
{ \int_{-L/2}^{L/2} {dx^\prime \over \lambda\revr{[ c_0(x^\prime)]}} \over \int_{-L/2}^{L/2} {dx^\prime \over \lambda\revr{[ c_0(x^\prime)]}\, c_0(x^\prime)}},
\label{dforce}
\end{align}
where we used the equilibrium condition $\partial_x( \mu[c_0(x)]+{\cal U})=0$ to simplify the expressions. 

This expression can be rewritten in terms of 
the thermodynamic osmotic pressure in Eq.~\eqref{PiGeneral}, noting that 
\begin{equation}
\Delta \Pi = \Pi(c_{\rm res} + \delta c_0) -  \Pi(c_{\rm res}) = c_{\rm res} \mu^\prime[c_{\rm res}] \delta c_0.
\end{equation}
The average force on the membrane in Eq.~\eqref{dforce} can accordingly be re-expressed as
\begin{equation}
 \int_{-L/2}^{L/2} dx\; \delta c(x)(-\partial_x{\cal U})= \sigma \Delta \Pi,
\end{equation}
where the reflection coefficient $\sigma$ is now defined as
\begin{equation}
\label{sigmaO}
\sigma = 1- { \int_{-L/2}^{L/2} {dx^\prime \over \lambda[c_0(x^\prime)]} \over \int_{-L/2}^{L/2} {dx^\prime \over \lambda[c_0(x^\prime)]}{c_{\rm res}\over c_0(x^\prime)}}.
\end{equation}
Equation~\eqref{sigmaO} takes into account the non-linearities that arise from the deviation from the simple Boltzmann distribution and the dependence of mobility on concentration. 
The driving force is formally the same as in Eq.~\eqref{Starling}, and the solvent flux takes accordingly the form:
\begin{equation}
 Q = - \mathcal{L}_\text{hyd} \left(\Delta p -\sigma \Delta \Pi \right),
	\label{kk-gen}
\end{equation}
with $\Pi$ the general thermodynamic expression in Eq.~\eqref{PiGeneral}.

Similarly to the discussion leading to Eq.~\eqref{linkb},  we also find
\begin{equation}
J_s = - \frac{1}{\int_{-L/2}^{L/2} \frac{dx'}{\lambda[c_0(x')]}} (1 - \sigma) \Delta \Pi,
\label{SoluteO}
\end{equation}
showing that the coefficients $\sigma$ and $\omega$ are again related as $1-\sigma=\omega$.

Altogether, this derivation 
unifies the thermodynamic  and the mechanical perspectives on the osmotic pressure.

\section{\label{sec_dof}Diffusio-osmotic transport at high solute concentrations}

\begin{figure}[t]
\includegraphics*[width=0.7\columnwidth]{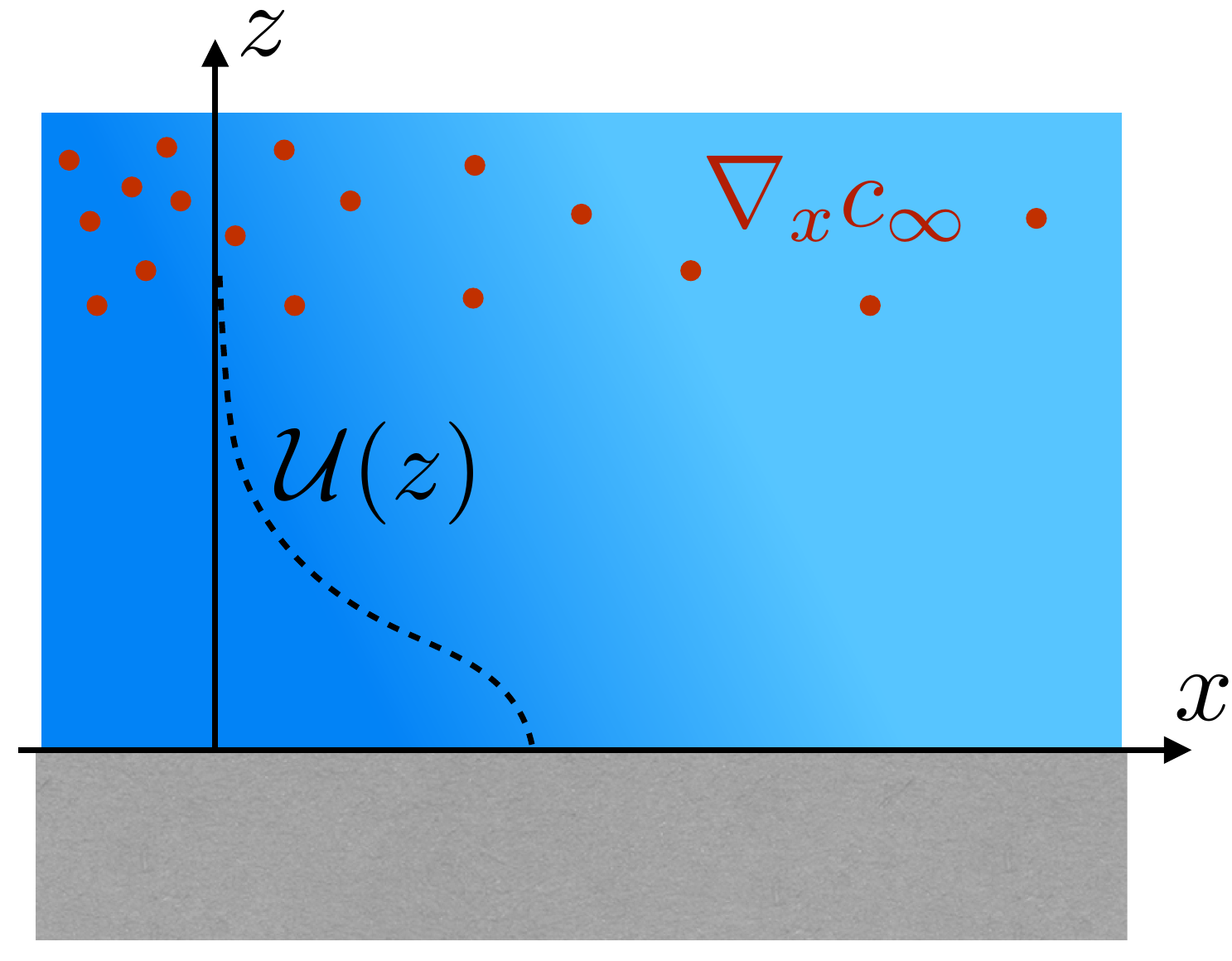}%
\caption{\label{fig:geoDO} Geometry of diffusio-osmosis only. Far from the surface, there is a uniform gradient of solute  $\nabla_x c_\infty$ parallel to the surface. The solute undergoes an external potential $\UU(z)$.
}
\end{figure}

We now explore similar questions for diffusio-osmotic transport. Diffusio-osmosis corresponds to the generation of solvent flow under a salinity gradient,
close to a solid surface,
see Fig.~\ref{fig:geoDO}. It is an interfacially driven phenomenon, which takes its origin within the 
diffuse interfacial layer close to the surface where the solute interacts
specifically with the surface.~\cite{Ruckenstein1981,Anderson1989,AB2006} Its effects were explored
in various experimental works.~\cite{SPB+2013,LCB+2014,SNA+2016,LCF+2017}
However only the regime of dilute solutes has been considered up to now, and we generalize the concepts to mixtures with high volume fraction of  the ``solute'' versus the ``solvent'' (solute and solvent being actually two components of a mixture). This will allow us to highlight the links between diffusio-osmosis and the generalized thermodynamic osmotic pressure, as introduced in Eq.~\eqref{PiGeneral}, and how it builds up.

The geometry is described in Fig.~\ref{fig:geoDO}. We consider a flat surface with a solute gradient along the membrane. 
We denote by $x$ the coordinate parallel to the surface, and by $z$ the one orthogonal to the surface. The solute concentration gradient far from the surface is $\partial_x c_\infty$ and is assumed to be uniform along $x$. Similarly to the mechanical approach for osmosis across a membrane, we introduce an external potential $\UU(z)$ from the surface, which acts only on the solute; one noticeable difference to the previous membrane case is that it now acts perpendicular to the solid surface and solute gradient ({\it i.e.} depending on $z$ but not on $x$). Typically, $\UU$ is strong near the surface within a thin layer and vanishes far from the surface.

We first focus on the solute distribution.
We assume a thin diffusive layer, {\it i.e.}, the equilibrium along $z$ is fast,
so that the gradient along the surface (along $x$) is small compared with the gradient orthogonal to the surface (along $z$).
In this case, local equilibrium establishes:
\begin{equation}
\mu[c(x,z)]+{\cal U}(z) \simeq \mu[c_\infty(x)].
\label{LocalEq}
\end{equation}
In the dilute regime where $\mu[c] = \kt \ln (c/c^*)$, we find $c(x,z) = c_{\infty}(x) \exp(-\UU(z)/\kt)$, as discussed above. It can be more complex
for high concentration $c$ of solute.
In general, depending on the interaction potential $\UU$, a surface excess or a surface depletion of the solute will occur at the surface.

This equation may be interpreted in terms of a mechanical balance introducing the generalized osmotic pressure.
Indeed,  using the local equilibrium condition Eq.~\eqref{LocalEq}, one gets
$c (-\partial_z{\cal U})= c \partial_z \mu= c\partial_z \left( {\partial f / \partial c}\right)\equiv  \partial_z \Pi[c]$ (the latter two equalities
being also a consequence of the Gibbs-Duhem relationship). Accordingly, one deduces that the
local force (along $z$) acting on the fluid due to the wall can be interpreted in terms of the osmotic
pressure as
\begin{equation}
c (-\partial_z {\cal U})= \partial_z \Pi[c].
\label{Pieq}
\end{equation}

Let us now turn to the fluid transport equation. It is described again by the Stokes equation:
\begin{equation}
0 = -\nabla p + c (-\nabla \UU)+ \eta \nabla^2 \vec{v}.
\label{St2}
\end{equation}
First, the projection along the $z$ direction (with vanishing component of the velocity $v_z$) yields
$0= - \partial_z p + c (-\partial_z{\cal U})$.
Using the previous result of Eq.~\eqref{Pieq}, one can rewrite this pressure balance as 
$\partial_z (p-\Pi)=0$, and obtain
\begin{equation}
p(x,z)-p_\infty= \Pi[c(x,z)]-\Pi[c_\infty(x)].
\label{pressure}
\end{equation}
Note that here, $\Pi[c]$ indeed takes the full expression of the osmotic pressure described in Eq.~\eqref{PiGeneral}.

This result then allows to obtain the solvent velocity profile. Indeed, 
injecting the pressure from  Eq.~\eqref{pressure} into the Stokes equation projected along $x$ gives 
\begin{equation}
0= \eta \partial_z^2 v_x - \partial_x (\Pi[c(x,z)]-\Pi[c_\infty(x)]).
\label{Stokesx}
\end{equation}
One then obtains the velocity field in terms of the osmotic pressure:
\begin{equation}
 v_x(x,z)={1\over \eta} \int_0^z\! dz^\prime\! \int_{z^{\prime}}^\infty\! dz^{\prime\prime} \partial_x (\Pi[c_\infty(x)]-\Pi[c(x,z^\prime\prime)]).
\end{equation}
Here the no slip boundary condition \revr{at the surface and a vanishing velocity gradient at infinity have} been assumed.

This expression can be rearranged 
using again the Gibbs-Duhem relation, $d\Pi=c d\mu$ (see Eq.~\eqref{PiGeneral}),
leading to 
\begin{equation}
 v_x(x,z) = {1\over \eta}\! \int_0^z\!\!\! dz^\prime\!\int_{z^{\prime}}^\infty\!\!\! dz^{\prime\prime} \left(c_\infty(x)- c(x,z^{\prime\prime})\right)
\partial_x \mu[c_\infty(x)].
\end{equation} 
Note that we used  
$\partial_x \mu[c(x,z)]= \partial_x \mu[c_\infty(x)]$ resulting from the local equilibrium in Eq.~\eqref{LocalEq}.
Finally, using $c_\infty(x)\partial_x \mu[c_\infty(x)] = \partial_x \Pi[c_\infty(x)]$ in the bulk, one obtains
a more transparent expression for the diffusio-osmotic velocity far from the surface as
\begin{equation}
v_\infty= K_{DO} \partial_x \Pi[c_\infty(x)],
\label{vDO}
\end{equation}
with the diffusio-osmotic mobility $K_{DO}$ given as
\begin{equation}
K_{DO}= - \frac{1}{\eta}\int_0^\infty dz^\prime\, z^\prime\, \left({c(x,z^\prime)\over c_\infty(x)} -1\right).
\label{KDO}
\end{equation}
Note that $\eta$ can also be assumed to depend on the concentration $c$. In this case, ${1/\eta}$ in $K_{DO}$ has to be integrated along $z$ as well.
The effect of hydrodynamic slippage on the surface can also be taken into account, along the same lines as in Ref.~\citenum{AB2006}, leading to
an enhancement factor  $(1 + b/L_s)$, where $b$ is the slip length and $L_s$ is the typical width of the diffuse interface.
\revr{The results for diffusio-osmosis in Eqs.~\eqref{vDO} and \eqref{KDO} are analogous to electro-osmosis and other 
surface-driven flows, with the mobility defined in terms of the first spatial moment of a density profile
(solute concentration profile for diffusio-osmosis and charge density profile in the case of electro-osmosis).
}

To sum up, a solute gradient generates an interfacial flow of the fluid. As highlighted in Eq.~\eqref{vDO}, this flow takes its origin in an osmotic pressure gradient occurring within the diffuse layer close to the surface.
Quantitatively, this flow is quantified by the value of the diffusio-osmotic mobility $K_{DO}$, which is non-zero only if there is surface excess or surface depletion of the solute. \revr{As a rule of thumb, its sign will be dominantly determined by the adsorption $\Gamma = \int_0^\infty dz^\prime\, \left({c(x,z^\prime)/ c_\infty(x)} -1\right)$.}
If there is a surface excess ($\Gamma>0$), the flow of water goes towards the low concentrated area ($K_{DO}<0$). Respectively, if there is a surface depletion, the flow of water reverses. \revr{But in case of a complex concentration profile, for instance with an oscillatory spatial dependence on $z$ due to layering, the sign of $K_{DO}$ may be expected to differ from the adsorption $\Gamma$. In this case, no obvious conclusion can be made for the direction of the diffusio-osmotic velocity and a full calculation has to be made.}

\revr{The general expression for the diffusio-osmotic velocity, in Eqs.~\eqref{vDO} and \eqref{KDO}, are very similar to 
 the corresponding expression for the dilute solution,~\cite{Ruckenstein1981,Anderson1989,AB2006}}
but the result in Eq.~\eqref{vDO} makes it very clear that 
the diffusio-osmotic flux is indeed driven by the thermodynamic osmotic pressure gradient $ \partial_x \Pi[c_\infty(x)]$. 

\section{Discussion and conclusions}

\revr{We conclude with a few words on possible extensions and implications of the results derived here.}
\subsection{Coupling osmosis and diffusio-osmosis}
\revr{First, while the previous derivations considered osmosis and diffusio-osmosis separately, one may consider a coupled situation in which both phenomena act jointly.}
\revr{Let us consider accordingly
a general situation of a {\it partially} permeable membrane, similar to Fig.~\ref{fig:geo}(b), now with a membrane interface interacting with solute particles}.
\revr{In full generality, the two transport phenomena are intimately coupled in the force balance and the situation is complex to disentangle. 
However some conclusions can be drawn in the limit where the pore size is large as compared to the interaction range of the surface potential, {\it i.e.} small diffuse layer.}
\revr{In this case, one may decompose  
the interaction potential into two contributions : $\UU(x,z)=\UU_{\infty}(x) + \revr{\UU_0(x,z)}$, where \revr{$\UU_0(x,z)$} is the surface interaction inside the membrane, which vanishes beyond the diffuse layer close to the surfaces; and $\UU_{\infty}(x)$ is independent of $z$ and describes a global energy barrier associated with the membrane, see Fig.\ref{fig:appB}.
Under this assumption, this situation is amenable to a full calculation, which we report in Appendix B.
}
As shown, the flow across the membrane still obeys a general Kedem--Kachalsky formula as in Eq.~\eqref{kk-gen}, with $\Pi$ the general osmotic force, but with a reflection coefficient
which now contains \revr{the coupled effects of osmosis and diffusio-osmosis}:
%
\begin{equation}
 \sigma = \sigma_{O} + \sigma_{DO} \revr{ -\sigma_{O}\sigma_{DO}},
\label{sigmaTot}	
\end{equation}
where $\sigma_{O}$ is the osmotic reflection coefficient given by Eq.~\eqref{sigmaO} and \revr{the diffusio-osmotic reflection coefficient is defined as $\sigma_{DO} = {\eta K_{DO}}/{\kappa}$, where $K_{DO}$ given in Eq.~\eqref{KDO} is the diffusio-osmotic reflection coefficient, and $\kappa$ the permeability.}

\revr{The last term in Eq.~\eqref{sigmaTot} accounts for a negative feedback coupling between osmosis and diffusio-osmosis. 
Although obtained in the limit of small diffuse layer, it exhibits a proper symmetry, as can be verified by considering various limiting situations. If the membrane is completely semi-permeable with $\sigma_{O} = 1$, then no solute flux occurs through the membrane and the diffusio-osmotic contribution in $\sigma_{DO} (1 - \sigma_O)$ vanishes. Reversely one also obtains $\sigma =1$ when $\sigma_{DO}=1$. A second note is that this coupled contribution hints to numerous possibilities to tune the flux through the membrane. With a given $\sigma_O < 1$, it is possible to enhance (respectively, diminish) $\sigma$ -- and thus the flux through the membrane -- with a slight surface depletion (respectively, excess) on the surface.}  
How this result may be generalized to any geometry remains to be explored.

\subsection{Outlook}

In summary, we have explored various forms of osmotic transport in the regime of high solute concentration. 
Both osmosis across model membranes and diffusio-osmosis at the interface with solid substrates were considered.
We have specifically focused our approach on the mechanical balance leading to solvent flow under
solute concentration gradients and explored the regime of high solute concentration. 
We demonstrate in particular how the general  expression of the osmotic pressure for mixtures, as 
obtained classically from the thermodynamic framework, emerges from the mechanical balance controlling 
non-equilibrium transport under solute gradients. 
The van\;'t Hoff expression for the osmotic pressure, $ \Pi = \kt  c$, is accordingly replaced by
its general thermodynamic counterpart, $\Pi = c\, \partial_c f - f[c] + f[c=0]$ (with $f$ the free energy density), 
which is valid for arbitrary composition of the ``solvent''/``solute'' mixture. 
This generalizes the existing results obtained in the dilute solute regime. 
In the second paper in this series, we will provide a numerical validation of the present results by means of molecular dynamic simulations.~\cite{YMB2017}  

\revr{An interesting consequence of the result in Eq.~\eqref{vDO} is that a non-linear ``sensing'' may originate from non-linearities of the osmotic pressure versus the solute concentration. This may occur, {\it e.g.}, for dense colloidal or polymer suspensions, as well as from a concentration dependent mobility $K_{DO}$ ({\it e.g.}, in ionic cases where $K_{DO}$ scales as the square of the concentration dependent Debye length).}
Finally let us discuss again the ingredients required to describe our osmotic membrane. 
In the case of osmosis, we considered a model membrane, where the details of the membrane are reduced
to its minimal ingredients and modeled as an external potential acting on the solute. As shown above, this
simplified picture is extremely fruitful to gain insight into the mechanical balance at play. 
But it  also points to the fact that osmosis does not require {\it per se} a solid membrane to be expressed. 
One may consider experimental situations where such a potential is built on the basis of optical
or electrical forces, using {\it e.g.} optical tweezers to repel ``solute'' particles, or dielectro-phoretic potential traps.
Such ``{\it osmosis without a membrane}'' configuration would be highly interesting to develop, as it would simplify many
aspects of clogging and pore blocking which occur for standard porous membranes. 
However designing such non-solid wells for molecular solutes, such as salts,
remains a considerable challenge.

\section*{Acknowledgments}
L.B. thanks fruitful discussions with B. Rotenberg, P. Warren, M. Cates and D. Frenkel on these topics.
L.B. acknowledges support from the European Union's FP7 Framework Programme/ERC Advanced Grant Micromegas. S.M. acknowledges funding from a J.-P. Aguilar grant. We acknowledge funding from
Agence Nationale de la Recherche (ANR) project BlueEnergy.

\section*{Appendix}

\subsection*{Appendix A : General expression of the osmotic pressure}

We recall below the general expression of the osmotic pressure given in Eq.~\eqref{PiGeneral}. The derivation is directly inspired by the presentation in Ref.~\citenum{BH2003}. 

Consider two compartments separated by a perfectly semi-permeable membrane. We consider on one side solvent with density $\rho_{A}$ and solute with density $\rho_B$. On the other side there are solvent with density $\rho'_{A}$ and solute with density $\rho'_B$. It is convenient to write the Helmholtz free energy of the mixture:
\begin{equation}
F(T,V,N_A,N_B) =-pV + \mu_A N_A +\mu_B N_B,
\end{equation}
where $V$ is the system volume, $N_A$ and $N_B$ are the number of solvent and solute particles, respectively. We used Euler's theorem with $ F = U - TS$, with $U$ the internal energy and $S$ the entropy.

The osmotic pressure is the difference in pressure between both sides:
\begin{align}
\Pi = P - P' =& \mu_A(\rho_A - \rho'_A) + \mu_B\rho_B - \mu'_B\rho'_B\nonumber\\
&+ f(T,\rho_A',\rho_B') - f(T,\rho_A,\rho_B),
\end{align}
where $f = F/V$ is now the free energy density. We also used the fact that the chemical potentials of the solvent at equilibrium are equal on both sides of the membrane:
\begin{equation}
\mu_A = \frac{\partial f(T,\rho_A',\rho_B')}{\rho_A'} = \frac{\partial f(T,\rho_A,\rho_B)}{\rho_A}.
\end{equation}
Now we define $\rho = \rho_A + \rho_B$ and $c = \rho_B$ in each compartment. As a result, (considering constant $\rho$,) we have in this new space variable
\begin{equation}
\mu_A = \frac{\partial f}{\partial c} \frac{\partial c}{\partial \rho_A} = 0 \hspace{2mm} {\rm and} \hspace{2mm} \mu_B = \frac{\partial f}{\partial c} \frac{\partial c}{\partial \rho_B} = \frac{\partial f}{\partial c}.
\end{equation}
We can thus express the osmotic pressure directly as
\begin{equation}
 \Pi = \frac{\partial f}{\partial c}c - \frac{\partial f}{\partial c'}c' + f(c') - f(c),
\end{equation}
which is exactly Eq.~\eqref{PiGeneral}, with $c' = 0$.
In the low density regime, we recall that $f(c) = c \kt \ln c + (1-c) \kt \ln (1-c)$ and thus we recover the limit of dilute systems, in which $\Pi = \kt (c- c')$.

\subsection*{Appendix B : \revr{Coupled osmotic and diffusio-osmotic flows}}

\begin{figure}[b]
\includegraphics*[width=1\columnwidth]{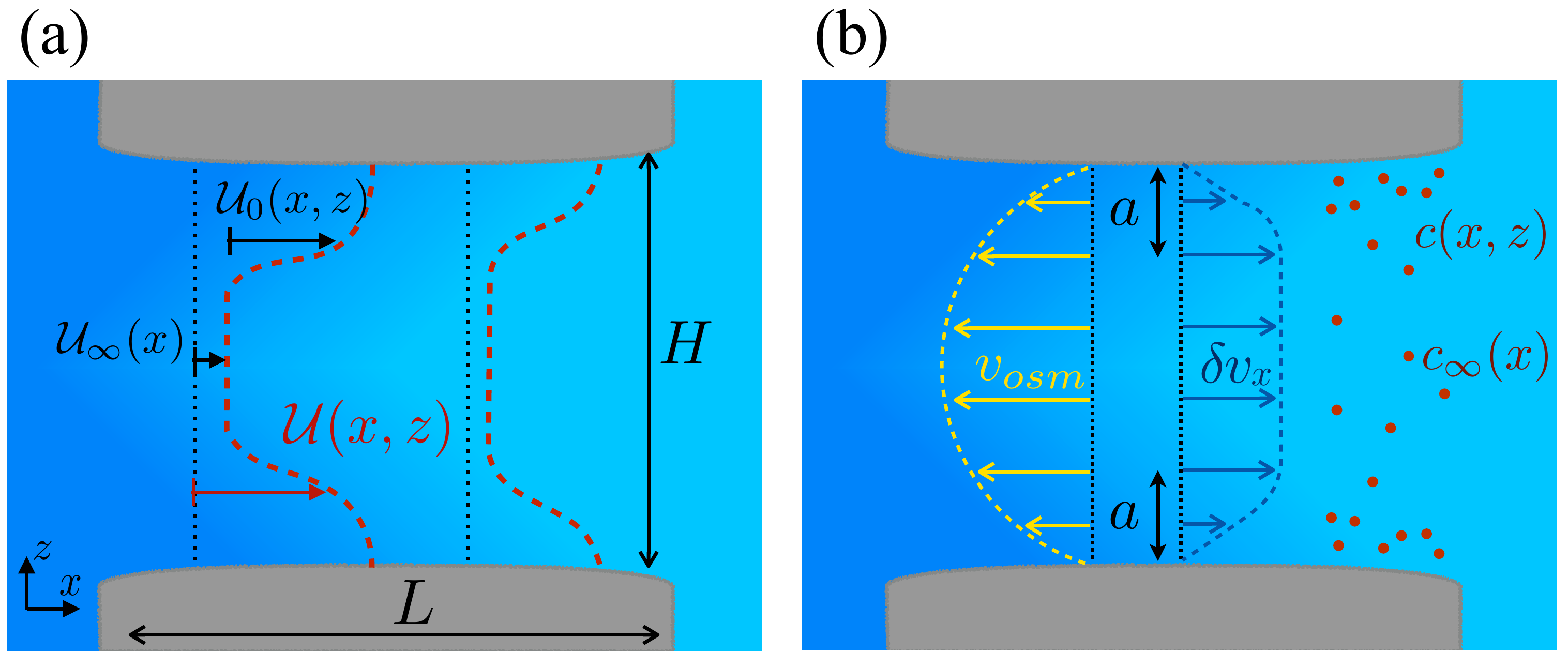}%
 \caption{\label{fig:appB} Geometry of a membrane pore, in which combined osmotic and diffusio-osmotic transport across the membrane take place.
 (a) Interaction potential expressed as $\UU(x,z)=\UU_{\infty}(x) + \UU_0(x,z)$, where $\UU_0(x,z)$ vanishes outside the diffuse layers.
 (b) Illustration of the decomposed velocity profiles $v_x=v_{osm}+\delta v_x$ within the pore.
 The typical thickness of the diffuse layer is denoted by $a$, and the typical size of the pore by $H$.
}
\end{figure}

Here we examine the case of a membrane with pores of a finite size interacting with the solute. Typically, the geometry under consideration is that of Fig.~\ref{fig:appB}. To simplify calculations, we assume a slit geometry with a planar pore of length $L$ and width $H$ (and invariant by translation in the perpendicular direction). We make also a further hypothesis and assume here that the interaction potential of the membrane with the solute can be decomposed as 
\begin{equation}
\UU(x,z) = \UU_{\infty}(x) + {\UU_0(x,z)},
\end{equation}
where $\UU(x,z)$ is non zero only between $-L/2$ and $L/2$,  $\UU_{\infty}(x) $ describes the global energy barrier associated with the membrane, and {$\UU_0(x,z)$} describes the specific interaction with the membrane pore surface, and vanishes for large $z$.
We write $a$ the typical range of the surface potential, fixing the size of the diffuse  layer near the membrane pore surfaces
(see. Fig.~\ref{fig:appB}.) The potential $\UU_{\infty}(x)$ is responsible for the osmotic transport and {$\UU_0(x,z)$} is responsible for the diffusio-osmotic transport.
Here we consider the case where no external pressure difference is applied between the two sides of the membrane; this can be simply
superimposed on the driving force as shown in Sec.~\ref{sec_osm}. To simplify the derivation, we make several additional simplifications. We assume that the deviation of the concentration profile to the equilibrium rest state is small.
Further, we assume that the thickness of the diffusive layer is sufficiently small as compared to the pore size, $a\ll H$.
In the following, we show how the reflection coefficients associated with osmotic and diffusio-osmotic transports are combined.

We write the thermodynamic equilibrium along the $z$ direction:
\begin{equation}
\label{eqPotentialNew}
\mu[c(x,z)] +  \UU(x,z) = \mu[c_{\infty}(x)] +  \UU_{\infty}(x).
\end{equation}
Here we have used the fact that the potential {$\UU_0(x,z)$} vanishes for large $z$. We derive first the osmotic current, following the steps of Sec.~\ref{sec_high}. We focus on the solute current out of the boundary layer, $z>a$, and compute there the solute profile starting from the Smoluchowski equation:
\begin{equation}
J_s^x = - \lambda[c_\infty(x)]c_{\infty}(x) \partial_x \left(  \mu[c_{\infty}(x)] +  \UU_{\infty}(x) \right).
\end{equation}
In parallel with the discussion in Sec.~\ref{sec_high}, we readily obtain the solute profile, and then compute the osmotic driving force as
\begin{equation}
\int_{-L/2}^{L/2}dx\; \delta c_{\infty}(x) (-\partial_x \UU_\infty) = \sigma_{O} \Delta \Pi,
\label{drive}
\end{equation}
with $\Delta \Pi= c_{\rm \infty, res} \mu^\prime[c_{\rm \infty, res}] \delta c_{\infty}$. The reflection coefficient of the osmotic transport here is defined as:
\begin{equation}
\sigma_{\rm O} = 1- { \int_{-L/2}^{L/2} {dx^\prime \over \lambda[c_{\infty,0}(x^\prime)]} \over \int_{-L/2}^{L/2} {dx^\prime \over \lambda[c_{\infty,0}(x^\prime)]}{c_{\rm \infty, res}\over c_{\infty,0}(x^\prime)}},
\end{equation}
where $c_{\infty,0}$ is the stationary concentration (see Sec.~\ref{sec_high}.)

Turning to the diffusio-osmotic part, one deduces the pressure profile from the Stokes equation along $z$,
similarly to Sec.~\ref{sec_dof} as 
\begin{equation}
p(x,z) - \Pi[c(x,z)] = p_{\infty} - \Pi[c_{\infty}(x)].
\end{equation}
The velocity profile is obtained from the Stokes equation along $x$:
\begin{align}
\eta \partial_{z}^2 v_x = &- \partial_x(\Pi[c_{\infty}(x)] - \Pi[c(x,z)]) - c(x,z)(-\partial_x \UU_0) \nonumber\\
 &  - c(x,z)(-\partial_x \UU_{\infty}).
\end{align}
Let us separate the flow into a bulk, pressure-driven, and a surface-driven contribution. We introduce accordingly
a velocity profile $v_{osm}(z)$ verifying 
\begin{equation}
\eta \partial_{z}^2 v_{osm} =    - c_\infty(x)(-\partial_x \UU_{\infty}),
\end{equation}
so that the remaining contribution to the velocity profile, $\delta v_x=v_x-v_{osm}$, verifies the equation
\begin{align}
\eta \partial_{z}^2 \delta v_x = &- \partial_x(\Pi[c_{\infty}(x)] - \Pi[c(x,z)]) - c(x,z)(-\partial_x \UU_0) \nonumber\\
 &  - (c(x,z)-c_\infty(x))(-\partial_x \UU_{\infty}),
\end{align}
and only contains surface-driven contributions.

The bulk contribution can be calculated following the very same steps as in Sec.~\ref{sec_high}. This  leads to a Poiseuille-like flow under the osmotic driving and, as in Eq.~\eqref{kk-gen} (with $\Delta p=0$), the corresponding averaged flux $Q_{osm}$  writes in terms of the permeability $\kappa$ of the pore as 
\begin{equation}
Q_{osm}=\frac{\kappa}{\eta} \sigma_O {\Delta \Pi\over L}.
\end{equation}
Here $\kappa$ scales typically as $H^2$ (in general, the exact expression for $\kappa$ will depend on the specific pore geometry).

The surface-driven contribution can be integrated as
%
\begin{align}
\eta\, \delta v_x (x,z)&= \int_0^z dz'\,z' \left( 1 - \frac{c(x,z')}{c_{\infty}(x)} \right) \partial_x \Pi[c_{\infty}(x)] \nonumber\\
 &\hspace*{0mm}  + \int_0^z dz'\,z'\, (c(x,z')-c_{\infty}(x))(-\partial_x \UU_{\infty}),
\end{align}
where we made use of the local equilibrium of Eq.~\eqref{eqPotentialNew} as $\partial_x \mu[c(x,z)] + \partial_x \UU_0 = \partial_x \mu[c_{\infty}(x)]$.
We assumed  no-slip boundary condition at $z = 0$ and a vanishing velocity gradient for any $z \gg a$, see Fig.~\ref{fig:appB}(b). 
We rewrite the previous velocity profile using the diffusio-osmotic mobility $K_{DO}$ defined in Eq.~\eqref{KDO}:
\begin{equation}
\delta v_x (x,z\ge a)= K_{DO} \left( \partial_x \Pi[c_{\infty}(x)]     - c_{\infty}(x) (-\partial_x \UU_{\infty}) \right).
\label{vxODO}
\end{equation}

To go further and to simplify calculations, we consider now a situation where the (potential) dependence of $K_{DO}$ on $x$ can be neglected. In the dilute regime, it depends on $x$ as  $c(x,z)/c_{\infty}(x) = \exp( - \UU_0(x,z)/\kt)$, so that it is sufficient to require that the surface potential $\UU_0$ is weakly dependent on $x$ along the pore surface. Note also that typically $K_{DO} \sim \Gamma \times a$ with $\Gamma$ the adsorption.
One can then calculate the corresponding averaged flux across the membrane thickness (along $x$) and over its  area $\mathcal{A}$, using the assumption of $a\ll H$.
Interestingly, using Eq.~\eqref{drive}, the last contribution in Eq.~\eqref{vxODO} averages to the osmotic driving forces,
$\sigma_O  {\Delta \Pi}/{L}$
\begin{align}
 Q_{surf}  &= \frac{1}{\mathcal{A}\, L}\int d\mathcal{A}  \int_{-L/2}^{L/2} dx  \, \delta v_x (z) \nonumber \\
 &\hspace*{-3mm}=  {K}_{DO} (1-\sigma_O) \times \frac{\Delta \Pi}{L}.    
\label{vxODOb}
\end{align}
Note that we assumed here to simplify that the diffusio-osmotic mobility does not depend on $x$. 

Now, adding the two contributions to the flux, $Q_{osm}$ and $Q_{surf}$, one gets the total flux as
\begin{align}
Q&=Q_{osm}+Q_{surf} \nonumber\\
&= \frac{\kappa}{\eta} \sigma_O {\Delta \Pi\over L} + {K}_{DO} (1-\sigma_O) \times \frac{\Delta \Pi}{L}.
\label{Qtot}
\end{align}
Defining a diffusio-osmotic reflection coefficient  as
\begin{equation}
\sigma_{\rm DO} = \frac{\eta {K}_{DO}}{\kappa} = \frac{1}{\kappa} \int_0^a dz'\, z' \, \left( \frac{c(x,z')}{c_{\infty}(x)} - 1 \right),
\end{equation}
(where again the dependence on $x$ in the  integral is not considered here for simplification), we then rewrite Eq.~\eqref{Qtot} as
\begin{equation}
 Q  = \mathcal{L}_{\text{hyd}} \left( \sigma_{DO} + \sigma_{O} - \sigma_{DO}\sigma_{O} \right) \Delta \Pi,
\label{additivity}
\end{equation}
where, as before, the permeance is defined  $\mathcal{L}_\text{hyd} = \kappa /(\eta L)$.
This equation highlights the introduction of a global reflection coefficient $\sigma=\sigma_{DO} + \sigma_{O} - \sigma_{DO}\sigma_{O}$. This gives the expression for the total reflection coefficient of Eq.~\eqref{sigmaTot} in the main text.


%

\end{document}